# Polar enhancement of optical nonlinearities and domain-driven second harmonic contrast in bismuth telluro-halide van der Waals crystals

Kevin W. C. Kwock[1,2]*, Tenzin Norden[1,]*, Thomas P. Darlington[3], Kaiyuan Yao[3], Kai Du[4], Ana-Marija Nedić[5], Thaís V. Trevisan[5], Ekaterina Dolgopolova[1], Peter P. Orth[5], Rohit P. Prasankumar[1,6], P. James Schuck[2], Sang-Wook Cheong[4†], Wilton J. M. Kort-Kamp[7,†], Prashant Padmanabhan[1,†]

**Affiliations:**

[1]Center for Integrated Nanotechnologies, Los Alamos National Laboratory; Los Alamos, NM, 87545, USA

[2]Department of Electrical Engineering, Columbia University; New York, NY, 10027, USA

[3]Department of Mechanical Engineering, Columbia University, New York, NY 10027, USA

[4]Rutgers Center for Emergent Materials and Department of Physics and Astronomy, Rutgers University, Piscataway, NJ 08854, USA

[5]Department of Physics, Iowa State University; Ames, IA, 50011, USA

[6]Intellectual Ventures, Bellevue, WA 98005, USA

[7]Theoretical Division, Los Alamos National Laboratory; Los Alamos, NM, 87545, USA

\* Authors contributed equally

† Corresponding authors: sangc@physics.rutgers.edu, kortkamp@lanl.gov, prashpad@lanl.gov

**Abstract**: The BiTeX family of polar van der Waals (vdW) semiconductors offers a unique platform for exploring the interplay between polar crystalline structure and nonlinear optical phenomena. Here, we utilize second-harmonic generation (SHG) polarimetry to demonstrate giant anisotropic optical nonlinearities in BiTeBr and BiTeI driven by contributions to the crystals' nonlinear polarizability originating from their permanent dipole moment. In addition, using SHG microscopy, we show that BiTeI displays a distinctive SHG spatial texture consisting of thread-like regions of reduced harmonic intensity. These features demark the boundaries between phase and anti-phase polar domains, confirmed via piezoresponse force microscopy and are attributable to disorder-driven restoration of inversion symmetry and concomitant optical interference effects. Our results unveil the power of polarization-resolved SHG microscopy in elucidating the intricate relationship between structure, symmetry, and nonlinear optical responses in polar vdW materials and highlight the promise of BiTeX as a material platform for domain-engineered nanoscale nonlinear photonics.

Nonlinear optical (NLO) processes are integral to a wide array of applications, including tunable coherent light sources[1], quantum sensing devices[2–4], and optical communication hardware[5,6]. Such applications typically rely on materials with broken inversion symmetry, which allow for three-wave photonic mixing phenomena enabled through the second-order nonlinear susceptibility of the crystal[7,8]. In the last decade, the need to enhance the strength of such NLO processes has fueled efforts to identify materials with large intrinsic nonlinearities and polar materials, such as the Weyl semimetal TaAs[9], have seen renewed interest due to their support of giant second-order responses. Concurrently, the ever-growing push for device miniaturization has prompted the need to shrink the fundamental length scale of nonlinear photonic systems. This has motivated the search for new material classes, particularly van der Waals (vdW) crystals, capable of supporting robust NLO responses at the nanoscale[10–14].

At the intersection of these two research fronts lies a promising new class of intrinsically polar vdW materials. A notable example is the family of ternary bismuth telluro-halide (BiTeX, X = I, Br, or Cl) semiconductors, which host a wealth of emergent properties, including Rashba spin-splitting[15,16], complex Fermi surfaces[17–19], and topological phase transitions[20,21]. BiTeX crystals are composed of a triangular network of bismuth atoms sandwiched between tellurium and halide layers (Fig. 1a). This structural arrangement, characterized by the $P3m1$ space group, possesses continuously broken inversion symmetry from the bulk to the monolayer limit[21,22]. Furthermore, strong covalent Bi-Te coupling yields a $(BiTe)^+$ layer complex that is ionically bonded to the electronegative halide layer, leading to a large permanent dipole moment oriented along the $c$-axis of the crystal[23]. Together, these properties make BiTeX a unique platform to investigate the influence of polar crystal structure on the nature of NLO responses in vdW systems.

Here, we use second harmonic generation (SHG) polarimetry to unambiguously and quantitatively reveal that the polar crystal structure of BiTeX leads to a substantial enhancement of its $c$-axis-dependent second-order nonlinear susceptibility tensor component, compared to those associated with the $a$- and $b$-axes. We interpret this using a quantum theory of the second-order susceptibility, from which emerges an additional contribution to the nonlinear polarizability of the material arising from a permanent dipole moment; this contribution is maximized when the driving electric field polarization is aligned along the crystal's polar axis. Confocal SHG microscopy also reveals a complex thread-like network in the second harmonic response from the $ab$-plane of BiTeI, perfectly correlated with its polar domain structure that can be observed with piezoresponse force microscopy (PFM). This SHG texture stems from a strong suppression of the harmonic intensity at the domain boundaries, which we attribute to the combined effects of structural disorder and destructive interference between adjacent phase and antiphase domains. These findings establish a direct link between polar crystal structure and the NLO properties of BiTeX, highlighting the immense potential of this material family for robust, domain-engineered nanophotonics applications.

*Contrasting SHG responses from the ab-planes of BiTeI and BiTeBr*
In our studies, bulk crystals of BiTeBr and BiTeI, grown using a modified Bridgman technique (see methods), were mechanically exfoliated onto $SiO_2$/Si substrates. The bulk-like flakes were then subjected to focused femtosecond fundamental pulses (800 nm center wavelength, ~150 femtosecond pulse width) at normal incidence and the back-reflected SHG intensity was analyzed using polarization-resolved single-channel detection (Fig. 1b). The red pattern in Fig. 1c shows the SHG response from the $ab$-plane of BiTeI, where the fundamental field's linear polarization was continuously rotated by 360°, and the polarization of the measured SHG was locked parallel to that of the fundamental (i.e., a co-rotation scheme that simulates the rotation of the sample under fixed input/output polarization). A clear six-fold pattern can be seen, matching the expected polarization dependence of the SHG intensity associated with the $3m$ point group, given by

$$I_{ab}^{\text{BiTeX}}(\theta) = |d_{22}^{\text{BiTeX}}|^2 \cos^2 3\theta, \tag{1}$$

where $\theta$ is the angle between the fundamental polarization and the $X = [0,-1,0]$ direction and $d_{ij}$ is the second-order susceptibility tensor element under Kleinmann symmetry (see Supplementary Information, Sections I and II). The SHG response from BiTeBr shows an analogous six-lobed polarization dependence (blue pattern in Fig. 1c). However, it is noticeably stronger, nearly five times larger than that of BiTeI and almost twice as large as the response from the prototypical nonlinear photonic crystal (111) GaAs (green pattern in Fig. 1c).

To make an accurate comparison between the magnitudes of the relevant $d_{ij}$ for BiTeI and BiTeBr, Bloembergen-Pershan corrections must be used to properly scale the measured SHG intensity from each material to account for Fresnel effects associated with the reflection geometry used in our experiments[24]. Taking these into account (see Supplementary Information, Section III) and given that $I_{(111)}^{\text{GaAs}}(\theta) = (2/3)|d_{14}^{\text{GaAs}}|^2 \cos^2 3\theta$ and the reported $d_{14}^{\text{GaAs}} \approx 380$ pm/V for GaAs[25], we obtain $d_{22}^{\text{BiTeI}} \approx 94.1$ pm/V and $d_{22}^{\text{BiTeBr}} \approx 401$ pm/V. The latter is more than twice as large as the recently reported bulk nonlinearity of the emerging vdW crystal NbOCl$_2$[11]. In addition, given the narrow bandgap of the BiTeX family, resonant enhancement could lead to even larger values of $d_{22}^{\text{BiTeX}}$ further into the infrared regime. Indeed, we observe significant increases in the SHG intensity from both BiTeI and BiTeBr under longer wavelength fundamental excitation (see Supplementary Figure S2).

*Enhancement of the c-axis dependent nonlinear susceptibility in BiTeI*
As shown in Eq. (1), the $ab$-plane SHG response of BiTeX at normal incidence is only a function of $d_{22}^{\text{BiTeX}}$; this tensor component only involves the mixing of incident light field components with polarizations that lie entirely in the $ab$-plane. However, the $3m$ point group allows for other nonzero components of $\overleftrightarrow{\mathbf{d}}^{\text{BiTeX}}$, namely $d_{15}^{\text{BiTeX}}$, $d_{31}^{\text{BiTeX}}$, and $d_{33}^{\text{BiTeX}}$, that involve the mixing of incident fields polarization projections along the out-of-plane direction (i.e., the polar $c$-axis). Accordingly, varying the fundamental pulse's incidence angle can enable contributions from these components to the total SHG response. Indeed, we observe that the anisotropy of the SHG pattern of BiTeBr changes significantly with increasing incidence angle (see Supplementary Figure S3). Moreover, there is an enhancement of the maximum $p$-polarized SHG intensity as the incidence angle is increased, relative to the maximum $s$-polarized response (see Supplementary Figure S3). This is suggestive of a sizeable contribution to the SHG output originating from the other non-zero nonlinear susceptibility tensor elements. Nevertheless, the influence of Fresnel effects significantly complicates a quantitative comparison of the magnitude of these elements from such angle of incidence dependent measurements, necessitating another approach to make such an accurate assessment.

Due to vdW bonding, BiTeX crystals preferentially cleave in such a way as to readily expose pristine $ab$-plane surfaces; in our experience, this is the only smooth facet that can be prepared through the mechanical exfoliation of BiTeBr. However, we found that the exfoliation of large BiTeI crystals also produces small, but optically flat surfaces containing the $c$-axis (Fig. 2a). This allows us to directly probe the frequency mixing of light-fields with a polarization perfectly parallel to the polar axis of the crystal in a normal incidence geometry. The responses of this '$c$-containing-plane' and $ab$-plane (Fig. 2b), measured under identical experimental conditions, reveal a bi-lobed polar pattern from the former that is in stark contrast to the characteristic six-lobed structure of the latter. Remarkably, the peak SHG intensity from the $c$-containing-plane is ~50-times larger than that of the $ab$-plane. This can be interpreted from the analytical form of the $c$-containing-plane SHG response, given by (see Supplementary Information, Section I)

$$I(\theta) \propto |d_{33}\cos^3\theta + (2d_{15} + d_{31})\sin^2\theta \cos\theta + d_{22}\sin^3\theta \sin 3\gamma|^2, \qquad (2)$$

where $\theta$ is the angle between the incident fundamental field's polarization and the $c$-axis (which now lies in-plane) and $\gamma$ is the angle between the $b$-axis and the sample's out-of-plane direction (schematically shown in Supplementary Figure S1).

From Eq. (2), we see that a dominant $d_{22}^{\text{BiTeI}}$ term (i.e., $d_{22}^{\text{BiTeI}} \gg d_{15}^{\text{BiTeI}}, d_{31}^{\text{BiTeI}}, d_{33}^{\text{BiTeI}}$) could lead to the observed two-lobed SHG response from the $c$-containing plane for $0 < \gamma < 90°$. However, this possibility can be easily ruled out since the maximal magnitude of the $c$-containing-plane response should be nearly equivalent to (for $\gamma = 30°$) or smaller than (for $\gamma \neq 30°$) that of the $ab$-plane, contrary to the results shown in Fig 2b. In contrast, a dominant $d_{33}^{\text{BiTeI}}$ term (i.e., $d_{33}^{\text{BiTeI}} \gg d_{22}^{\text{BiTeI}}, d_{15}^{\text{BiTeI}}, d_{31}^{\text{BiTeI}}$) also yields a two-lobed SHG pattern from the $c$-containing-plane (first panel of Fig. 2c), while having no impact on the $ab$-plane SHG response at normal incidence (as seen from Eq. (1)). This interpretation agrees with the differences in the maximum intensity and overall shape of the two patterns shown in Fig. 2b. Furthermore, when considering the second term of equation (2), a moderate $(2d_{15}^{\text{BiTeI}} + d_{31}^{\text{BiTeI}})$ primarily leads to a broadening of such a two-lobed pattern (top right panel of Fig. 2c), rather than an increase in its maximal intensity. In contrast, a large $(2d_{15}^{\text{BiTeI}} + d_{31}^{\text{BiTeI}})$ results in the gradual emergence of a four-lobed pattern (bottom panels of Fig. 2c). We can therefore ascribe the effective nonlinearity associated with the peak of the $c$-containing-plane SHG response to be $d_{\text{eff},c}^{\text{BiTeI}} \approx d_{33}^{\text{BiTeI}}$. Additionally, given the giant enhancement of the maximum SHG intensity from the $c$-containing-plane relative to the $ab$-plane, we estimate that $|d_{33}^{\text{BiTeI}}| \approx 665$ pm/V, nearly seven-times the magnitude of $|d_{22}^{\text{BiTeI}}|$. In addition, given the larger permanent dipole moment of BiTeBr relative to BiTeI[23], we expect that $d_{33}^{\text{BiTeBr}} > d_{33}^{\text{BiTeI}}$, and comparable to the $d_{33}^{\text{TaAs}}$ second-order susceptibility tensor element responsible for the giant second harmonic response of the polar Weyl semimetal TaAs[9]. Most importantly, however, this analysis provides direct experimental evidence that the permanent dipole moment significantly enhances the intrinsic second-order optical nonlinearity of BiTeX uniquely associated with the polar axis (i.e., the $c$-axis) of the crystal.

### *Quantum theory of the polar enhancement of nonlinear susceptibility*
To elucidate the role of a permanent dipole moment on the second-order susceptibility of BiTeX, we consider a finite-sized flake of the material, in which case the electronic bandgap is well described by discrete energy levels (Fig. 3a). When interacting with the light-field, an electron in a state $|g\rangle$ can undergo virtual transitions to higher energy states through photon absorption (e.g., at $\omega_1$ and $\omega_2$ as shown in Fig. 3a), emitting a harmonic at $\omega_3$ upon decaying. For the case of second order NLO processes, three photons are involved and sum frequency generation (the generalization of SHG involving incident photons that may have different energies) is described by six Feynman diagrams illustrated in Fig. 3b. The simplest are (i) and (ii), which describe a sequence involving the absorption of a single photon leading to an excitation from $|g\rangle$ to a virtual state $|l\rangle$, the absorption of a second photon leading to an excitation from $|l\rangle$ to $|e\rangle$, and the emission of a higher energy photon as the electron decays from $|e\rangle$ to $|g\rangle$. For multiple photon absorption, each virtual transition does not need to necessarily conserve energy; only the total energy must be conserved in the process. This is the case for diagrams (iii) – (vi), which involve a permutation of the photon absorption/emission processes.

We consider the six processes delineated in Fig. 3b within a quantum perturbation theory formalism via the density matrix operator (see Supplementary Information, Section VI), allowing us to express the second-order nonlinear polarizability as $\vec{\mathbf{P}}^{(2)}(\omega_3) = \vec{\mathbf{P}}_t^{(2)}(\omega_3) + \vec{\mathbf{P}}_p^{(2)}(\omega_3)$. Here, $\vec{\mathbf{P}}_t^{(2)}$ corresponds to the typical contribution of transition dipole moments to the total second-order polarizability in all materials with broken inversion symmetry[8]; it is the only relevant contribution in systems with a vanishing permanent dipole moment. Under quasi-resonant emission conditions ($\hbar\omega_3 = E_e - E_g$) and assuming that no resonant absorption occurs, the second term, $\vec{\mathbf{P}}_p^{(2)}$, for the case of SHG can be derived in the crystal's optical axes reference frame as

$$\vec{\mathbf{P}}_p^{(2)}(2\omega) \cong \frac{N}{\hbar^2}\left[\hat{\rho}_{gg}^{(0)} - \hat{\rho}_{ee}^{(0)}\right]\vec{\mathbf{\mu}}^{ge}\frac{\left[\vec{\mathbf{\mu}}^{eg}\cdot\vec{\mathbf{E}}(\omega)\right]\left[\vec{\mathcal{D}}\cdot\vec{\mathbf{E}}(\omega)\right]}{\left[\omega_{eg} - 2\omega - i\gamma_{eg}\right]\left[\omega_{eg} - \omega - i\gamma_{eg}\right]}. \tag{3}$$

Here, $N$ is the number of flakes per volume, $\hat{\rho}_{nm}^{(0)}$ are matrix elements of the density matrix operator at equilibrium, $\mu_i^{nm}$ are matrix elements of the $i$-th cartesian component of the electric dipole moment operator, $\gamma_{nm}$ is the relaxation rate for the transition from $|n\rangle$ to $|m\rangle$, $\omega_{nm} = (E_n - E_m)/\hbar$, where $E_n$ is the energy of the state $|n\rangle$, and we define $\vec{\mathcal{D}} = \vec{\mathbf{\mu}}^{ee} - \vec{\mathbf{\mu}}^{gg}$ as the difference between the permanent electric dipole moments of the $|e\rangle$ and $|g\rangle$ states.

Eq. (3) clearly shows that a nonzero permanent dipole moment $\vec{\mathcal{D}}$ yields a new contribution to the traditional second-order nonlinear polarizability. Importantly, if $\vec{\mathcal{D}}$ is much larger than the transition dipole moments, as is the case for strongly polar BiTeX, $\vec{\mathbf{P}}_p^{(2)}$ can be the dominant term in $\vec{\mathbf{P}}^{(2)}$ under appropriate conditions. In particular, the permanent dipole moment provides a maximal enhancement of the second-order polarizability when $\vec{\mathcal{D}} \parallel \vec{\mathbf{E}}$. Contrastingly, its contribution identically vanishes when $\vec{\mathcal{D}} \perp \vec{\mathbf{E}}$. This agrees with our observation of a significant $c$-containing-plane SHG enhancement in BiTeI at normal incidence, where the electric field of the driving pulse can be perfectly parallel to the crystal's $c$-axis (i.e., the permanent dipole moment).

### *SHG textures revealed via confocal-SHG microscopy*
We now turn to the exploration of the microscopic structure of the NLO responses of BiTeI. Figure 4a shows the co-polarized SHG map from the $ab$-plane where we observe regions of intense SHG separated by a thread-like network of reduced harmonic intensity. To gain further insight, PFM was used to characterize the crystal's polar domain structure. Fig. 4b shows a magnified area of the SHG image (yellow box in Fig. 4a), revealing an exact correspondence with the domain features over the same region captured with PFM (Fig. 4c). Here, black and white regions are associated with domains with different surface terminations (i.e., Te or I) and antiparallel permanent dipole moment vectors (i.e., phase and anti-phase domains)[26]. Intriguingly, the harmonic intensity is similar for both domains, with the SHG image contrast stemming from the reduced signal intensity along their boundaries. Examining the SHG polar patterns between adjacent domains and their boundary (red, green, and blue dots, respectively, in Fig. 4b) shows no appreciable change in either the shape or orientation of the polar patterns between the three regions (Fig. 4d). Analogous measurements on BiTeBr revealed no significant texture in its SHG response, in-line with its monodomain character[27]. This may also explain the larger $ab$-plane SHG response of BiTeBr, relative to BiTeI (Fig. 1c).

The SHG image contrast could have a few possible sources; in the vicinity of the domain wall there could be (1) a re-orientation of the permanent dipole moment vector, (2) a change in its magnitude, (3) a change in the structural arrangement of the crystal, or (4) optical interference effects. The first possibility would necessarily lead to a non-zero projection of the permanent dipole moment vector onto the $ab$-plane as it rotates from parallel to anti-parallel (or vice versa) with respect to the $c$-axis. This would likely lead to an enhancement of the $a$- and/or $b$-axis dependent nonlinear susceptibility tensor elements between adjacent domains and, given the contribution to the nonlinear polarizability expressed by Eq. (3), would result in stronger SHG signal at the domain boundaries, rather than the observed suppression. Regarding the second and third possibilities, the similarity in the ionic sizes of tellurium and iodine could lead to a tellurium-iodine disordered phase in the vicinity of the domain wall. A randomized occupation of the anion sites would restore mirror symmetry along the $c$-axis, yielding the centrosymmetric $P\bar{3}m1$ space group. This is also the likely structural phase of BiTeX in its non-polar form[22], and would therefore point to a concomitant reduction in the magnitude of the permanent dipole moment vector in the disordered region around the domain wall. As such, increased Te/I disorder between the phase and anti-phase domains (Fig. 4e) could lead to the observed suppression of the SHG intensity in this region. Finally, for the fourth possibility, the $\pi$-phase shift between the second harmonic fields generated from the phase- and anti-phase domains should lead to a drop in the second harmonic intensity at the domain boundary due to destructive interference[28]. Therefore, we surmise that both disorder-induced inversion symmetry restoration and interference effects contribute to the polar-domain-driven SHG image contrast.

## *Conclusion*

Our investigation of the second harmonic responses of bismuth telluro-halide layered crystals marks an important advance in our understanding of the interplay of polar crystal structure on the NLO properties of this novel material family. The enhanced second-order nonlinearity stemming from their intrinsic polar moment is uniquely compatible with the weak interlayer bonding and structural tunability inherent to these vdW systems, offering a new material platform for robust nonlinear nanophotonics. Looking forward, nanoscale characterization of their NLO responses could provide unprecedented access to the microscopic structure and symmetry of polar domain walls; these regions may also harbor distinct electronic states that could impact their nonlinear polarizability. Such capabilities not only offer a route to directly visualize symmetry-breaking phenomena in layered polar semiconductors but also lay the groundwork for the design of nonlinear optical functionalities through domain engineering[29]. In particular, given their narrow band gap, these systems are particularly promising for applications in the mid-infrared regime, where ultrathin, vertically integrable hosts of frequency conversion remain a critical challenge in the development of nanophotonic technologies.

# References

1. Manzoni, C. & Cerullo, G. Design criteria for ultrafast optical parametric amplifiers. *J. Opt.* **18**, 103501 (2016).

2. Lawrie, B. J., Lett, P. D., Marino, A. M. & Pooser, R. C. Quantum Sensing with Squeezed Light. *ACS Photonics* **6**, 1307–1318 (2019).

3. Moreau, P.-A., Toninelli, E., Gregory, T. & Padgett, M. J. Imaging with quantum states of light. *Nat. Rev. Phys.* **1**, 367–380 (2019).

4. Kutas, M. *et al.* Quantum Sensing with Extreme Light. *Adv. Quantum Technol.* **5**, 2100164 (2022).

5. Schneider, T. *Nonlinear Optics in Telecommunications*. (Springer, Berlin Heidelberg, 2010).

6. Agrawal, G. P. Nonlinear fiber optics: its history and recent progress [Invited]. *J. Opt. Soc. Am. B* **28**, A1 (2011).

7. Boyd, R. W. *Nonlinear Optics*. (Academic Press, Burlington, MA, 2008).

8. Shen, Y. R. *The Principles of Nonlinear Optics*. (Wiley-Interscience, Hoboken, N.J, 2003).

9. Wu, L. *et al.* Giant anisotropic nonlinear optical response in transition metal monopnictide Weyl semimetals. *Nat. Phys.* **13**, 350–355 (2017).

10. Trovatello, C. *et al.* Optical parametric amplification by monolayer transition metal dichalcogenides. *Nat. Photonics* **15**, 6–10 (2021).

11. Guo, Q. *et al.* Ultrathin quantum light source with van der Waals NbOCl2 crystal. *Nature* **613**, 53–59 (2023).

12. Yao, K. *et al.* Continuous Wave Sum Frequency Generation and Imaging of Monolayer and Heterobilayer Two-Dimensional Semiconductors. *ACS Nano* **14**, 708–714 (2020).

13. Norden, T. *et al.* Twisted Nonlinear Optics in Monolayer van der Waals Crystals. *ACS Nano* **19**, 30919–30929 (2025).

14. Autere, A. *et al.* Nonlinear Optics with 2D Layered Materials. *Adv. Mater.* **30**, 1705963 (2018).

15. Ishizaka, K. *et al.* Giant Rashba-type spin splitting in bulk BiTeI. *Nat. Mater.* **10**, 521–526 (2011).

16. Ideue, T. *et al.* Bulk rectification effect in a polar semiconductor. *Nat. Phys.* **13**, 578–583 (2017).

17. Landolt, G. *et al.* Disentanglement of Surface and Bulk Rashba Spin Splittings in Noncentrosymmetric BiTeI. *Phys. Rev. Lett.* **109**, 116403 (2012).

18. Ye, L., Checkelsky, J. G., Kagawa, F. & Tokura, Y. Transport signatures of Fermi surface topology change in BiTeI. *Phys. Rev. B* **91**, 201104 (2015).

19. Ideue, T. *et al.* Thermoelectric probe for Fermi surface topology in the three-dimensional Rashba semiconductor BiTeI. *Phys. Rev. B* **92**, 115144 (2015).

20. Qi, Y. *et al.* Topological Quantum Phase Transition and Superconductivity Induced by Pressure in the Bismuth Tellurohalide BiTeI. *Adv. Mater.* **29**, 1605965 (2017).

21. Ohmura, A. *et al.* Pressure-induced topological phase transition in the polar semiconductor BiTeBr. *Phys. Rev. B* **95**, 125203 (2017).

22. Du, K. *et al.* Kibble–Zurek mechanism of Ising domains. *Nat. Phys.* **19**, 1495–1501 (2023).

23. Ma, Y., Dai, Y., Wei, W., Li, X. & Huang, B. Emergence of electric polarity in BiTeX (X = Br and I) monolayers and the giant Rashba spin splitting. *Phys. Chem. Chem. Phys.* **16**, 17603 (2014).

24. Bloembergen, N. & Pershan, P. S. Light Waves at the Boundary of Nonlinear Media. *Phys. Rev.* **128**, 606–622 (1962).

25. Bergfeld, S. & Daum, W. Second-Harmonic Generation in GaAs: Experiment versus Theoretical Predictions of $\chi_{xyz}^{(2)}$. *Phys. Rev. Lett.* **90**, 036801 (2003).

26. Butler, C. J. *et al.* Mapping polarization induced surface band bending on the Rashba semiconductor BiTeI. *Nat. Commun.* **5**, 4066 (2014).

27. Fiedler, S. *et al.* Termination-dependent surface properties in the giant-Rashba semiconductors BiTeX (X = Cl, Br, I). *Phys. Rev. B* **92**, 235430 (2015).

28. Kaneshiro, J., Uesu, Y. & Fukui, T. Visibility of inverted domain structures using the second harmonic generation microscope: Comparison of interference and non-interference cases. *J. Opt. Soc. Am. B* **27**, 888 (2010).

29. Weiss, T. F. & Peruzzo, A. Nonlinear domain engineering for quantum technologies. *Appl. Phys. Rev.* **12**, 011318 (2025).

**Methods**

*Crystal growth.* High-quality, single crystalline BiTeX materials were prepared through a modified Bridgman technique. Stoichiometric bismuth shot, tellurium powder, and wither bromine or iodine were sealed in an evacuated quartz tube with a small excess of iodine. The sealed ampules were then annealed at 700°C for 24 hours at a heating rate of 20~30°C/h. After cooling the furnace, the ampule was then flipped and placed upside down in a vertical furnace and subsequently heated at 700°C for 12 hours. The ampule was then cooled down to 400°C with different cooling rate at a small temperature gradient. For the crystal post-annealing process, the crystals were placed in the evacuated quartz tube and heated up to 500°C.

*SHG polarimetry.* Polarization-resolved SHG was performed on freshly-cleaved $ab$-plane oriented flakes of BiTeX and on freshly-cleaved single crystals (in the case of $c$-containing plane experiments). A tunable pulsed laser (Coherent Ti:sapphire oscillator emitting 150 fs pulses with a repetition rate of 80 MHz) served as the excitation source. The excitation beam was focused onto the sample with a 20X apochromatic objective (Mitutoyo) and the resulting SHG signal was transmitted through a Glan-Laser polarizer (mounted on a motorized stage) and filtered using a series of bandpass filters at the second harmonic wavelength before being focused into a photomultiplier (Hamamatsu R943-02). An achromatic half-wave plate (mounted on a motorized stage) was used to control the polarization of the excitation beam.

***PFM and nonlinear optical imaging.*** Freshly-cleaved *ab*-plane oriented flakes of BiTeX were scanned by PFM with a conductive Pt/Ir-coated contact-mode tip at room temperature with an atomic force microscope (Brooker Dimension). Roughly AC 1 V was applied to the tip at 75 kHz during the scan.

***Nonlinear optical microscopy.*** Nonlinear optical scanning microscopy was performed on freshly-cleaved *ab*-plane oriented flakes of BiTeX using a customized inverted microscope (based on a Nikon Eclipse Ti−S inverted microscope). A tunable pulsed laser (Coherent Ti:sapphire oscillator emitting 150 fs pulses with a repetition rate of 80 MHz) served as the excitation source. The excitation beam was focused through a high numerical aperture (NA= 0.95), 100X apochromatic objective. Samples were mounted on a three-dimensional (XYZ) nano-scanning piezo stage (PI P-545.xR8S Plano) for sample-scanning second harmonic generation confocal imaging. The second harmonic emission was collected with the same objective and filtered through relevant short-pass optical filters and then directed either onto a single-photon avalanche diode (Micro Photon Device, PDM series).

**Acknowledgements**
K.W.C.K., T.N., W.J.M.K.-K., and P.P. acknowledge support from Los Alamos National Laboratory (LANL) Laboratory Directed Research and Development program (20220273ER and 20240037DR). K.W.C.K. also acknowledges support from the DOE National Nuclear Security Administration (NNSA) Laboratory Residency Graduate Fellowship Program No. DE-NA0003960, NSF award DMR-1747426, and the Programmable Quantum Materials, an Energy Frontier Research Center funded by the US DOE, Office of Science, Office of Basic Energy Sciences, award DESC0019443. This work was performed, in part, at the Center for Integrated Nanotechnologies, an Office of Science User Facility operated for the U.S. DOE Office of Science. Los Alamos National Laboratory, an affirmative action equal opportunity employer, is managed by Triad National Security, LLC for the U.S. DOE's NNSA, under contract 89233218CNA000001.

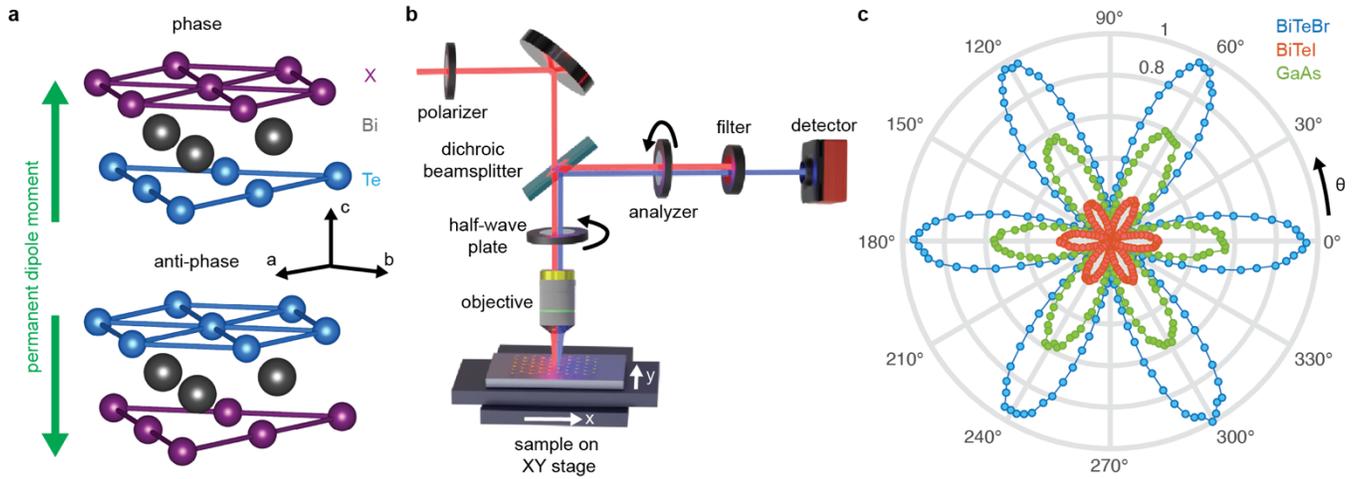

**Fig. 1. Crystal structure, SHG experiment, and *ab*-plane SHG response from BiTeX. a**, Crystal structures of the phase (top) and anti-phase (bottom) structural domains of BiTeX with green arrows indicating their net polarization direction; **b**, schematic of the SHG microscopy/imaging system; **c**, SHG polarimetry of BiTeBr (blue pattern), BiTeI (red pattern), and (111) GaAs (green pattern) under a co-rotation scheme measured under identical normal incidence conditions with an 800 nm fundamental pulse.

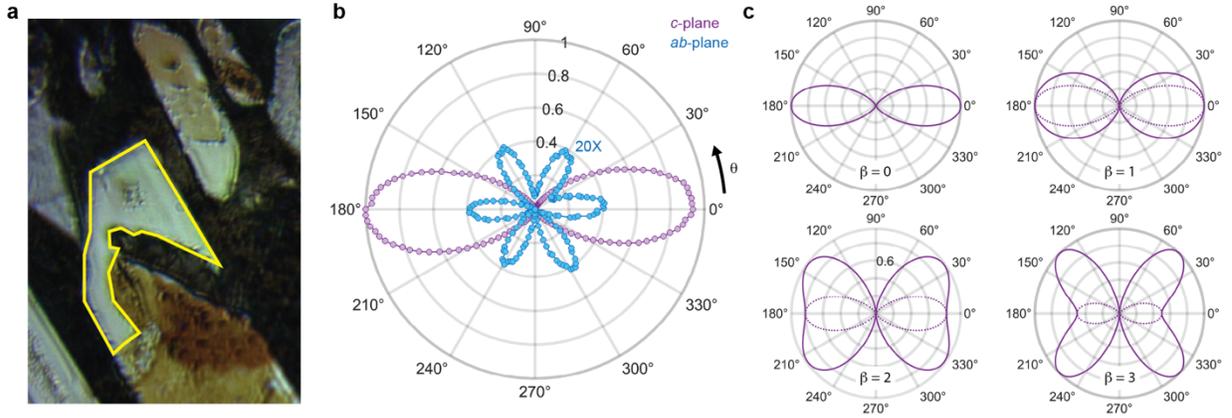

**Fig. 2 | BiTeI *c*-plane and its SHG response. a**, Small *c*-plane surface (outlined in yellow) exposed by mechanically exfoliating a large bulk crystal of BiTeI; **b**, SHG polarimetry of the *c*-axis-containing plane (purple pattern) compared to the *ab*-plane response (blue pattern, scaled by a factor of 20) under a co-rotation scheme and measured under identical normal incidence conditions; **c**, theoretically expected polar patterns from the *c*-axis in-plane of BiTeI under a normal incidence co-rotation scheme, calculated using Eq. (2) for $\gamma = 0$, for various values of $\beta = (2d_{15} + d_{31})/d_{33}$, where the dashed two-lobed pattern in the top right and bottom panels is the same as the top left panel to allow for a relative comparison of magnitude.

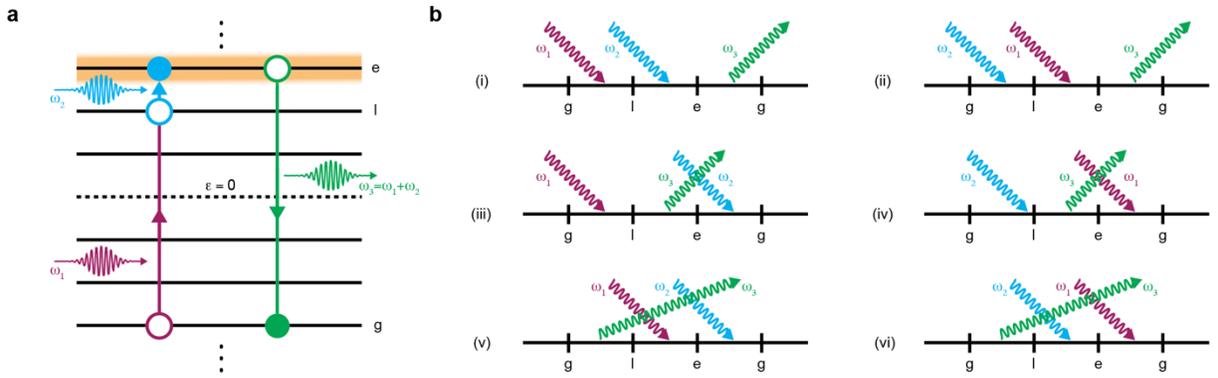

**Fig. 3 | Model system and Feynman diagrams of sum frequency generation. a**, multi-level model of sum frequency generation showing the virtual optical transitions following the absorption of a photon at $\omega_1$ (cyan) and $\omega_2$ (burgundy) and the emission of a harmonic at $\omega_3$ (green), where $g$, $l$, and $e$ refer to eigenstates; **b**, Feynman diagrams of the sum frequency generation process where time proceeds from left to right and processes (iii) – (vi) have non-energy-conserving intermediary steps. In **a**, the orange glow around eigenstate $e$ signifies that it carries a non-zero permanent electric dipole moment.

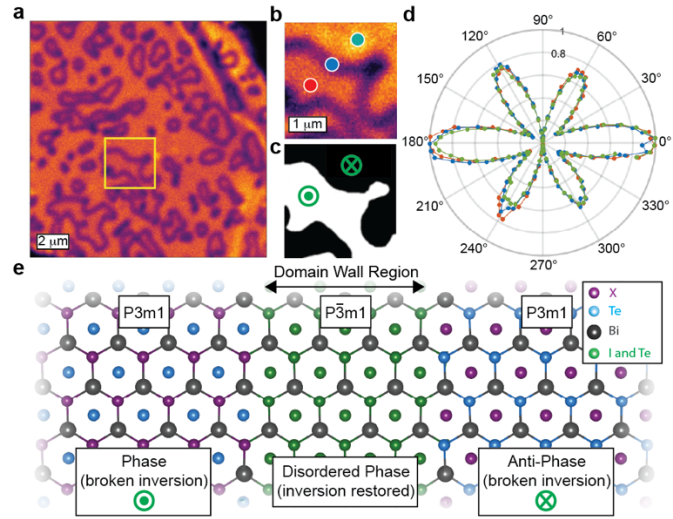

**Fig. 4 | Polar domains of BiTeI imaged with SHG microscopy. a**, SHG image of the *ab*-plane surface of BiTeI showing thread like regions of suppressed SHG intensity; **b**, a zoomed in image of the region in **a** outlined in yellow; **c,** PFM image of the same region as in **b**, where the green dot indicates an out-of-plane permanent dipole moment and the green cross indicates an into-the-plane permanent dipole moment; **d**, normalized SHG polarimetry for the three regions delineated by red, blue, and green dots in **b**; **e**, schematic of the *ab*-plane surface of BiTeX showing the phase, disordered (domain wall), and anti-phase regions, noting their space groups, structural inversion symmetry properties, and the orientation of the permanent dipole moment in the phase and anti-phase regions.